\documentclass[aps,prl,twocolumn,floatfix, showpacs]{revtex4}
\usepackage{graphicx}
\usepackage{dcolumn}
\usepackage{bm}
\usepackage{color}
\usepackage{natbib}
\usepackage{array}
\usepackage{subfigure}
\usepackage{amssymb}
\usepackage{amsbsy}

\def\gtwid{\mathrel{\raise.3ex\hbox{$>$\kern-.75em\lower1ex\hbox{$\sim$}}}}
\def\ltwid{\mathrel{\raise.3ex\hbox{$<$\kern-.75em\lower1ex\hbox{$\sim$}}}}

\begin{document}

\title{Optimal Prandtl number for heat transfer in rotating Rayleigh-B\'enard convection}
\author{Richard J.A.M. Stevens$^1$}
\author{Herman J.H. Clercx$^{2,3}$}
\author{Detlef Lohse$^1$}
\affiliation{$^1$Department of Science and Technology and J.M. Burgers Center for Fluid Dynamics, University of Twente, P.O Box 217, 7500 AE Enschede, The Netherlands}
\affiliation{$^2$Department of Applied Mathematics, University of Twente, P.O Box 217, 7500 AE Enschede, The Netherlands}
\affiliation{$^3$Department of Physics and J.M. Burgers Centre for Fluid Dynamics, Eindhoven University of Technology, P.O. Box 513, 5600 MB Eindhoven, The Netherlands}
\date{\today}

\begin{abstract}
Numerical data for the heat transfer as a function of the Prandtl ($Pr$) and Rossby ($Ro$) numbers in turbulent rotating Rayleigh-B\'enard convection are presented for Rayleigh number $Ra = 10^8$. When $Ro$ is fixed the heat transfer enhancement with respect to the non-rotating value shows a maximum as function of $Pr$. This maximum is due to the reduced efficiency of Ekman pumping when $Pr$ becomes too small or too large. When $Pr$ becomes small, i.e. for large thermal diffusivity, the heat that is carried by the vertical vortices spreads out in the middle of the cell, and Ekman pumping thus becomes less efficient. For higher $Pr$ the thermal boundary layers (BLs) are thinner than the kinetic BLs and therefore the Ekman vortices do not reach the thermal BL. This means that the fluid that is sucked into the vertical vortices is colder than for lower $Pr$ which limits the efficiency of the upwards heat transfer.
\end{abstract}


\maketitle

Turbulent Rayleigh-B\'enard (RB) convection, i.e. the flow of a fluid between two parallel plates heated from below and cooled from above, is the paradigmatic system for thermally driven turbulence in a confined space. Considerable progress on the understanding of the flow dependencies has been achieved in the last two decades as reviewed in \cite{ahl09,loh10}. Here we study the case where the sample is rotated around the vertical axis at an angular velocity $\Omega$. That system is relevant in the astro- and geophysical context, namely for convection in the arctic ocean \cite{mar99}, in Earth's outer core \cite{gla99}, in the interior of gaseous giant planets \cite{bus94}, and in the outer layer of the Sun \cite{mie00}.

Previous experimental \cite{liu09,ros69,pfo84,zho93,zho09b,ste09,kin09} and numerical studies \cite{ore07,jul96,kun08b,zho09b,ste09,kin09} have shown that the convective heat transport increases at moderate rotation rates, before it rapidly decreases for stronger rotation. This heat transport enhancement has been ascribed to Ekman pumping \cite{ros69,zho93,jul96,vor98,vor02,kun08b,zho09b,ste09,kin09}.  Namely, whenever a plume is formed, the converging radial fluid motion at the base of the plume (in the Ekman boundary layer) starts to swirl cyclonically resulting in the formation of vertical vortex tubes. The rising plume induces stretching of the vertical vortex tube and hence additional vorticity is created. This leads to enhanced suction of hot fluid out of the local Ekman layer and thus increased heat transport. Corresponding phenomena happen at the upper boundary. This process strongly depends on $Ra$ and $Pr$ \cite{zho09b} and the boundary conditions \cite{sch09}. In ref.\ \cite{zho09b} it was shown that no heat-transfer enhancement is observed at small $Pr\lesssim 0.7$. This was explained by the larger thermal diffusivity at lower $Pr$ due to which the heat that is carried by the vertical vortices created by Ekman pumping spreads out in the middle of the cell. This leads to a larger destabilizing temperature gradient in the bulk and a lower heat transfer. Furthermore, Ekman pumping turned out to become less efficient at higher $Ra$ numbers where a larger eddy thermal diffusivity limits the efficiency of Ekman pumping.

\begin{figure} [t]
\includegraphics[width=3.25in]{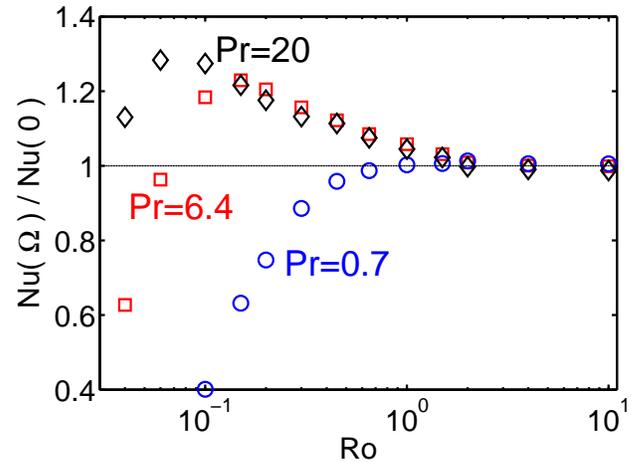}
\caption{The ratio $Nu(\Omega)/Nu(\Omega=0)$ as function of $Ro$ on a logarithmic scale for $Ra = 1\times 10^8$ and different $Pr$. Blue open circles: $Pr=0.7$, red open squares: $Pr=6.4$, and black open diamonds: $Pr=20$.}
\label{fig:Nu_1}
\end{figure}

In the present work we systematically determine the heat-transfer enhancement with respect to the non rotating value as function of the Prandtl ($Pr$) number by using direct numerical simulations (DNS). We find that in certain regimes the heat-flux enhancement can be as large as $30\%$. Even more remarkably, we observe a heretofore unanticipated maximum in the heat transfer enhancement as function of $Pr$.

For given aspect ratio $\Gamma \equiv D/L $ ($D$ is the cell diameter and $L$ its height) and given geometry, the nature of RB convection is determined by the Rayleigh number $Ra = \beta g \Delta L^3/(\kappa\nu)$
and the Prandtl number $Pr = \nu /\kappa$. Here, $\beta$ is the thermal expansion coefficient, $g$ the gravitational acceleration, $\Delta$ the temperature difference between the bottom and top, and $\nu$ and $\kappa$ the kinematic viscosity and the thermal diffusivity, respectively. The rotation rate $\Omega$ (given in rad/s) is non-dimensionalized in the form of the Rossby number
$
Ro =  \sqrt{\beta g \Delta /L}/(2\Omega)
$.
In the DNS we solved the three-dimensional Navier-Stokes equations within the Boussinesq approximation. The numerical scheme was already described in Refs. \cite{ver96,ver99,ver03,ore07,ste09b,ste09e}. The numerical details concerning rotating RB convection are described in detail in Ref. \cite{zho09b,ste09}. Most simulations were performed on a grid of $129 \times 257 \times 257$ nodes, respectively, in the radial, azimuthal, and vertical directions, allowing for a sufficient resolution of the small scales both inside the bulk of turbulence and in the BLs (where the grid-point density has been enhanced) for the parameters employed here \cite{zho09b,ste09}. $Nu$ is calculated in several ways as described in detail in Ref. \cite{ste09b} and its statistical convergence has been controlled. A grid of $193 \times 385 \times 385$ nodes has been used for the simulations at the highest $Pr$ number and to verify the results for $Pr=6.4$ obtained on the coarser grid. Furthermore, for the higher $Pr$ number cases the flow is simulated for a very long time ($400$ dimensionless time units to reach the statistically stationary state followed by $800$ time units for averaging) to assure statistical convergence. In refs. \cite{zho09b,ste09} we already showed that our DNS results agree very well with experimental results in this $Ra$ number regime.

\begin{figure} [t]
\subfigure{\includegraphics[width=3.25in]{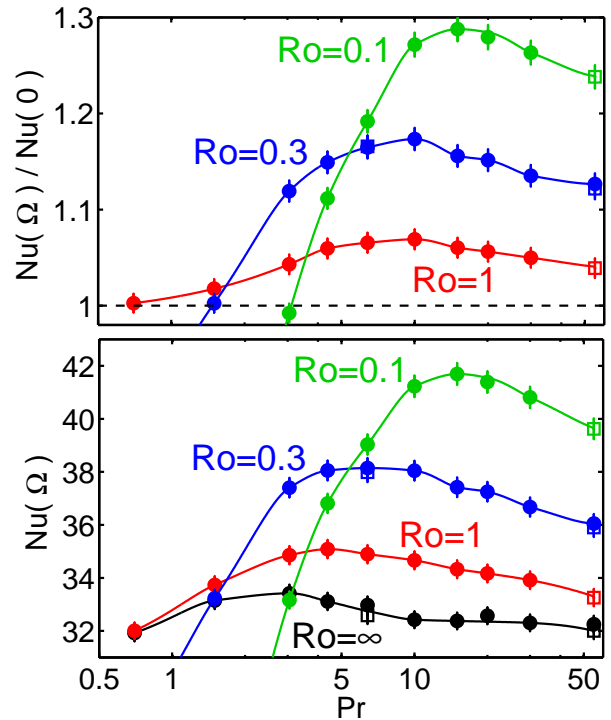}}
\caption{The heat transfer as function of $Pr$ on a logarithmic scale for $Ra = 1\times 10^8$. Black, red, blue, and green indicate the results for $Ro=\infty$, $Ro=1.0$, $Ro=0.3$, and $Ro=0.1$, respectively. The data obtained on the $193\times 385 \times 385$ (open squares) and the $129\times 257 \times 257$ grid (solid circles) are in very good agreement.}
\label{fig:Nu_2}
\end{figure}

The numerical results for $Nu(\Omega) / Nu(0)$ as function of $Ro$ for several $Pr$ are shown in figure \ref{fig:Nu_1}. The figure shows that the heat transport enhancement caused by Ekman pumping can be as large as $30\%$ for $Pr=20$. However, no heat transport enhancement is found for $Pr=0.7$ which is due to the larger thermal diffusivity which makes Ekman pumping less efficient \cite{zho09b}. In figure \ref{fig:Nu_1} one can already see that the heat transfer enhancement reaches a maximum at a certain $Pr$ when the $Ro$ number is fixed. Indeed figure \ref{fig:Nu_2} confirms that the heat transfer enhancement as function of $Pr$ reaches a maximum. Its location depends on $Ro$; namely the stronger the rotation rate the higher the $Pr$ number for which the maximum heat transfer enhancement is found. This trend is even more pronounced  when the heat transfer $Nu(Ro)$ itself is considered.

\begin{figure}
\centering
\subfigure[]{\includegraphics[width=3.25in]{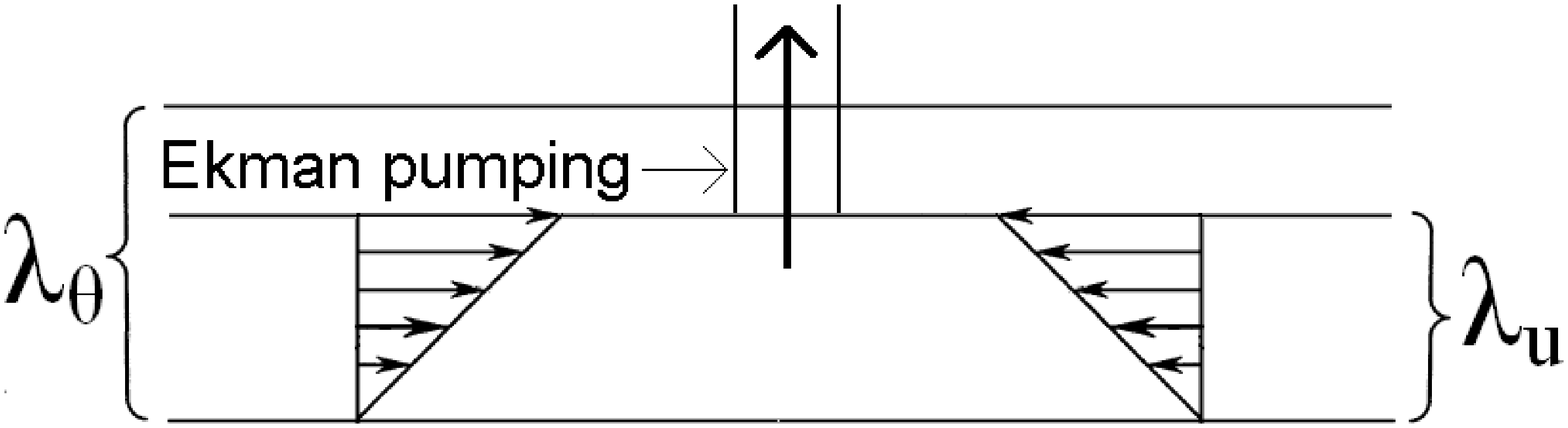}}
\subfigure[]{\includegraphics[width=3.25in]{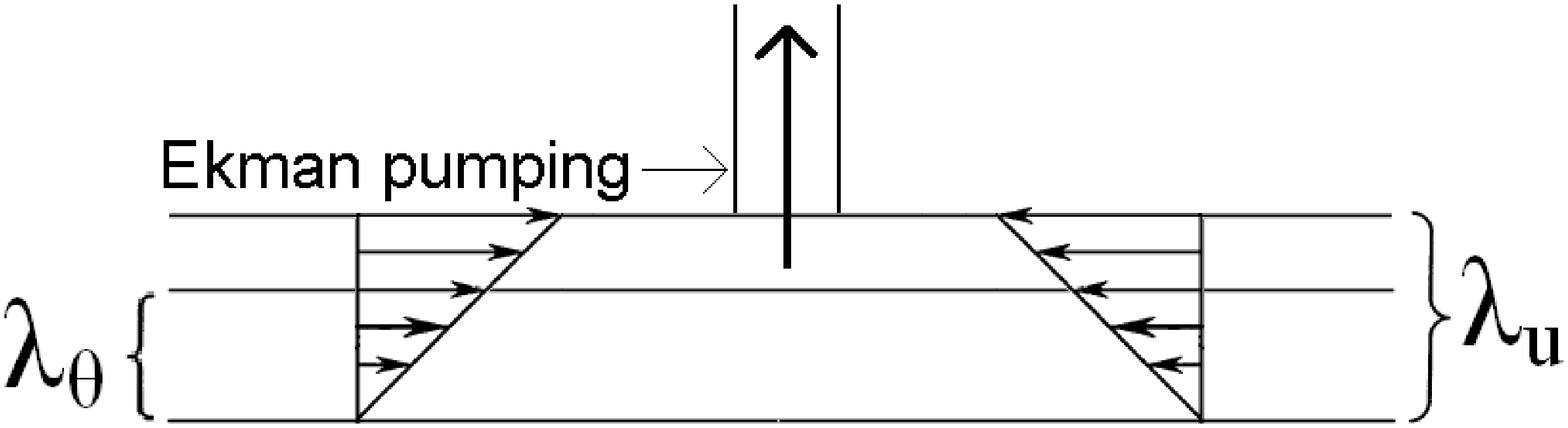}}
\caption{a) Sketch for the low $Pr$ number regime where the Ekman vortices reach the thermal BL. b) Sketch for the high $Pr$ number regime where the Ekman vortices do not reach the thermal BL.}
\label{fig:sketch}
\end{figure}

The observation that there is a $Pr$ number for which the heat transfer enhancement is largest suggests that there should be at least two competing effects, which strongly depend on the $Pr$ number, that control the efficiency of Ekman pumping. One wonders why Ekman pumping becomes less efficient for higher $Pr$ numbers. Clearly, it must have a different origin than the reduced efficiency of Ekman pumping at lower $Pr$. Indeed, there is an important difference between the high and the low $Pr$ number regime, namely the relation between the thickness of the thermal and the kinetic boundary layers (BL). For the low $Pr$ number regime the kinetic BL is thinner than the thermal BL and therefore the fluid that is sucked into the vertical vortices is very hot. When the $Pr$ number is too low this heat will spread out in the middle of the cell due to the large thermal diffusivity. For somewhat higher $Pr$ number the fluid that is sucked out of the thermal BL is still sufficiently hot and due to the smaller thermal diffusivity the heat can travel very far from the plate in the vertical vortices. In this way Ekman pumping can increase the heat transfer for moderate $Pr$. In the high $Pr$ number regime the kinetic BL is much thicker than the thermal BL. Therefore the Ekman vortices forming in the bulk do not reach the thermal BL and hence the temperature of the fluid that enters the vertical vortices is much lower. This is schematically shown in figure \ref{fig:sketch}.

\begin{figure}
\centering
\subfigure{\includegraphics[width=0.21\textwidth]{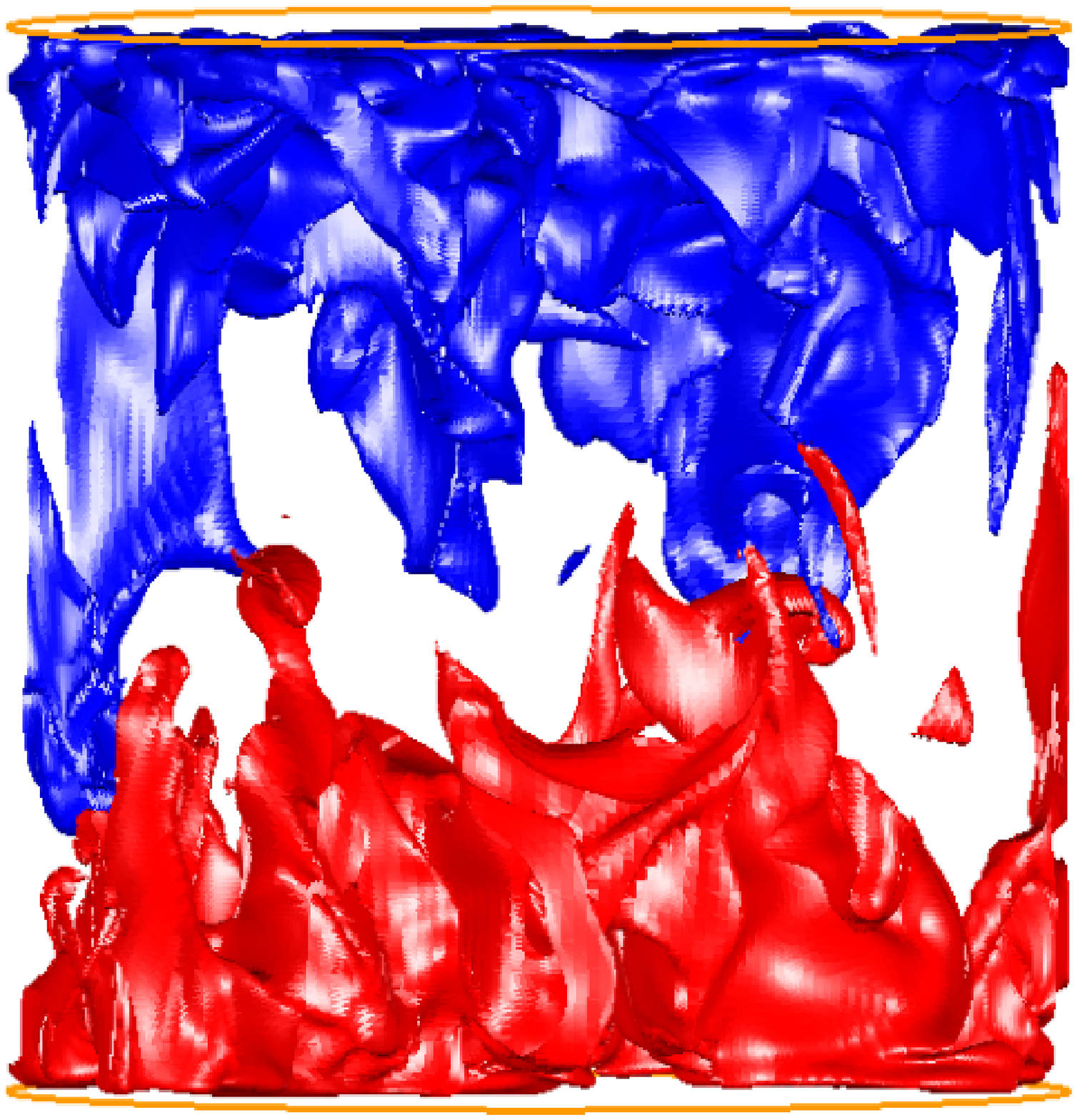}}
\hspace{2mm}
\subfigure{\includegraphics[width=0.21\textwidth]{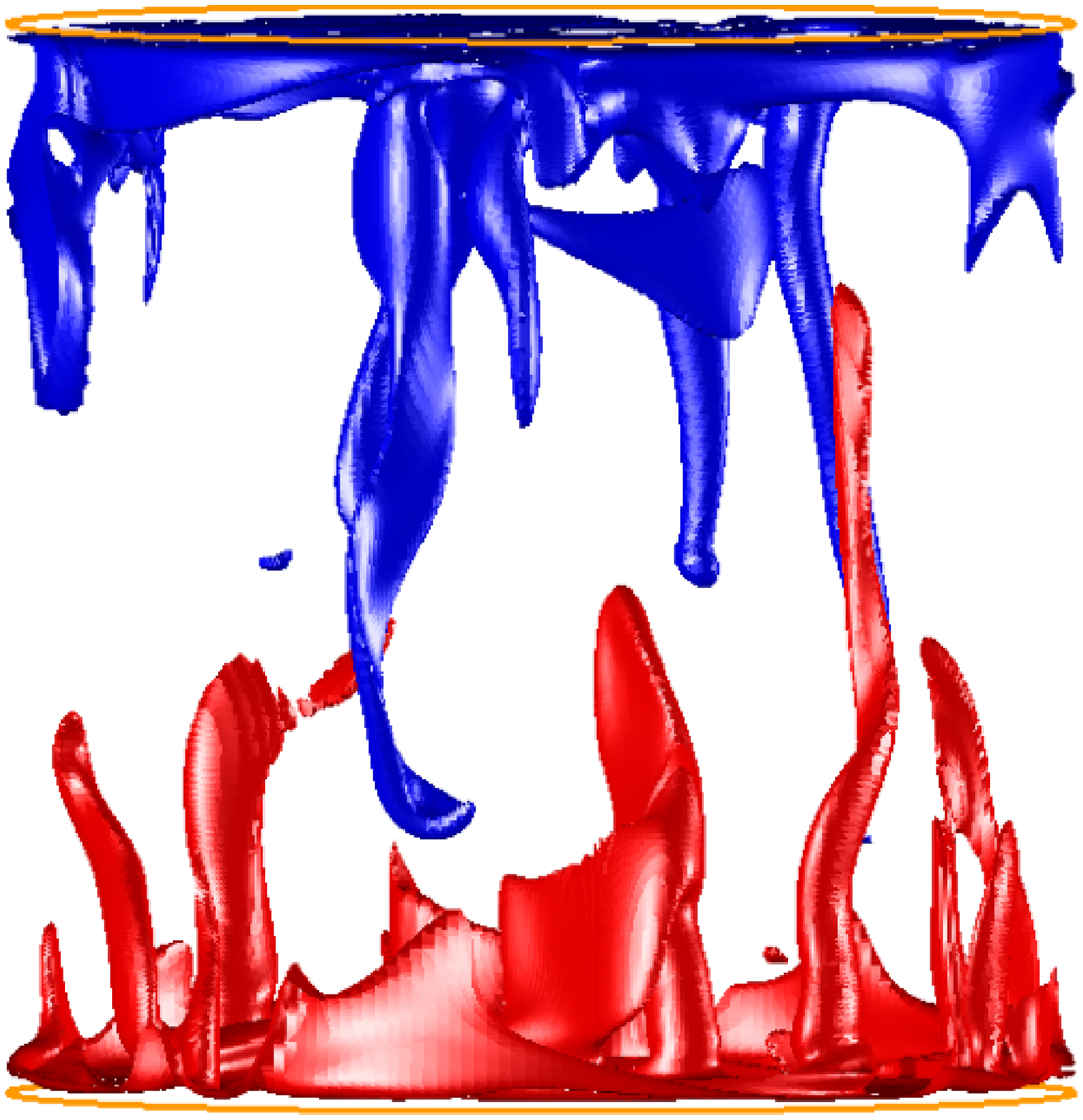}}
\vspace{1mm}
\subfigure{\includegraphics[width=0.21\textwidth]{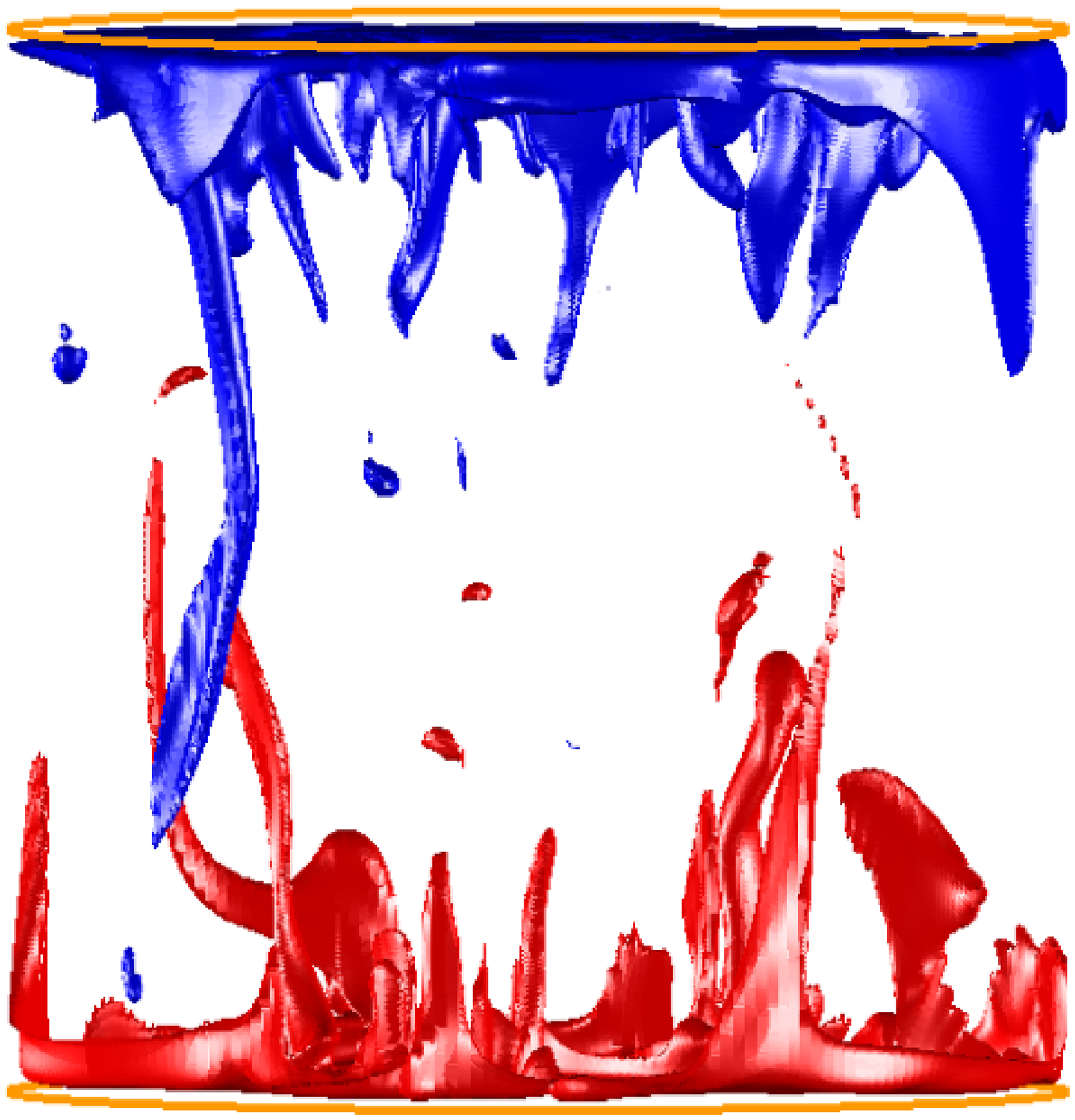}}
\hspace{2mm}
\subfigure{\includegraphics[width=0.21\textwidth]{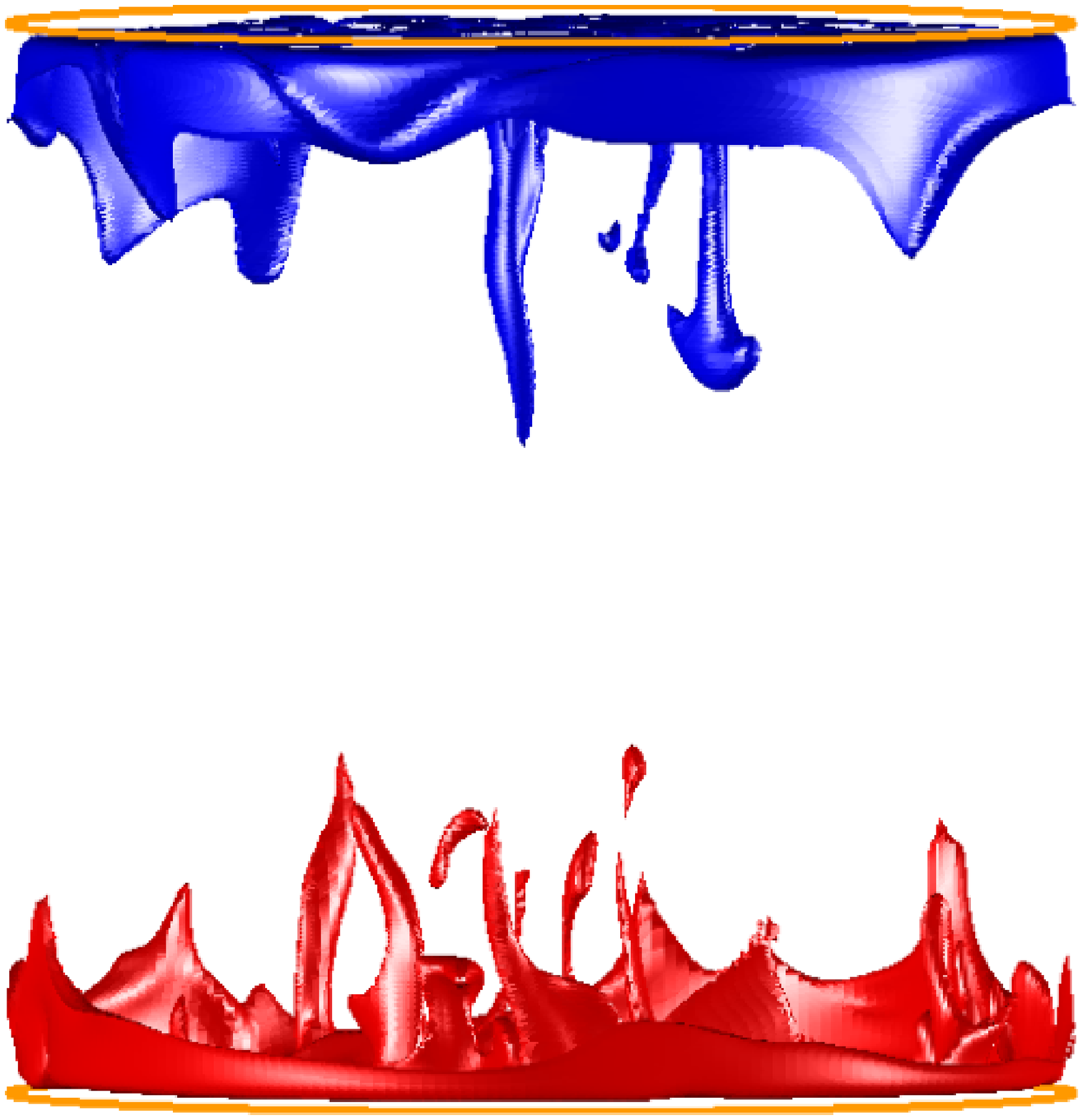}}
\caption{3D visualization of the temperature isosurfaces in the cylindrical sample of $\Gamma = 1$ at $0.65\Delta$ (red) and $0.35\Delta$ (blue), respectively, for $Pr = 0.7$ (left upper plot) and $Pr = 6.4$ (right upper plot), $Pr=20$ (left lower plot), and $Pr=55$ (right lower plot) for $Ra =  10^8$ and $Ro=0.30$. The snapshots were taken in the respective statistically stationary regimes.}
\label{fig:3D-plots}
\end{figure}

An investigation of the temperature isosurfaces \cite{zho09b} revealed long vertical vortices as suggested by Ekman-pumping at $Pr=6.4$, while these structures are much shorter and broadened for the $Pr = 0.7$ case due to the larger thermal diffusivity at lower $Pr$ \cite{zho09b}. In figure \ref{fig:3D-plots} we show the temperature isosurfaces at $Ro=0.30$ for several $Pr$ to identify the difference between the high and moderate $Pr$ number cases. The figure reveals that the vertical transport of hot (cold) fluid away from the bottom (top) plate through the vertical vortex tubes is strongly reduced in the high-Pr number regime compared to the case with $Pr=6.4$. This is illustrated in figure \ref{fig:3D-plots} where the threshold for the temperature isosurfaces is taken constant for all cases. Indeed, a closer investigation shows that the vortical structures for the higher $Pr$ are approximately as long as for the $Pr=6.4$ case. This confirms the view that the temperature of the fluid that is sucked into the Ekman vortices decreases with increasing $Pr$.

\begin{figure} [t]
\subfigure{\includegraphics[width=3.25in]{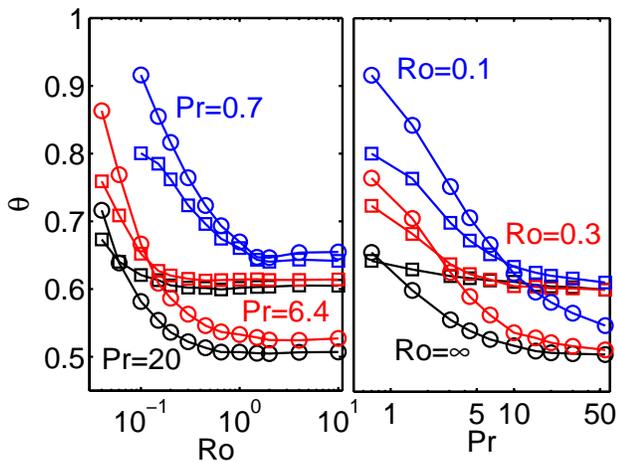}}
\caption{The horizontally averaged temperature at the edge of the kinetic BL (circles) and at the edge of the thermal BL (squares) for $Ra=1\times 10^8$. Left: Blue, Red, and black indicates the results for $Pr=0.7$, $Pr=6.4$ and $Pr=20$, respectively. Right: Black, red, and blue indicate the results for $Ro=\infty$, $Ro=0.3$, and $Ro=0.1$, respectively. The results for $Ro=1$ (not shown) are almost identical to the data obtained for $Ro=\infty$.}
\label{fig:Ekman}
\end{figure}

To further verify the above explanation we determined the horizontally averaged temperature at the kinetic BL height, defined as two times the height where the kinetic dissipation rate is highest, see Refs. \cite{ste09b,ste09e} for details. This height is chosen as it indicates the position where effects of Ekman pumping are dominant. Furthermore, we also determined the horizontally averaged temperature at the edge of the radially dependent thermal BL thickness, i.e. $\lambda_\theta^{sl}(r)$. We determined $\lambda_\theta^{sl}(r)$ by finding the intersection point between the linear extrapolation of the temperature gradient at the plate with the behavior found in the bulk, see Ref. \cite{ste09e}. Indeed, figure \ref{fig:Ekman} shows that the average temperature at the edge of the kinetic BL decreases with increasing $Pr$. Thus the temperature of the fluid that enters the vertical vortices decreases with increasing $Pr$ and this limits the efficiency of Ekman pumping at higher $Pr$. Furthermore, the temperature difference with respect to the bottom plate shows that for the non-rotating case the kinetic BL is thinner than the thermal BL when $Pr<1$ and thicker than the thermal BL when $Pr>1$. The figure shows that this transition point shifts towards higher $Pr$ when the rotation rate is increased.

\begin{figure} [t]
\subfigure{\includegraphics[width=3.25in]{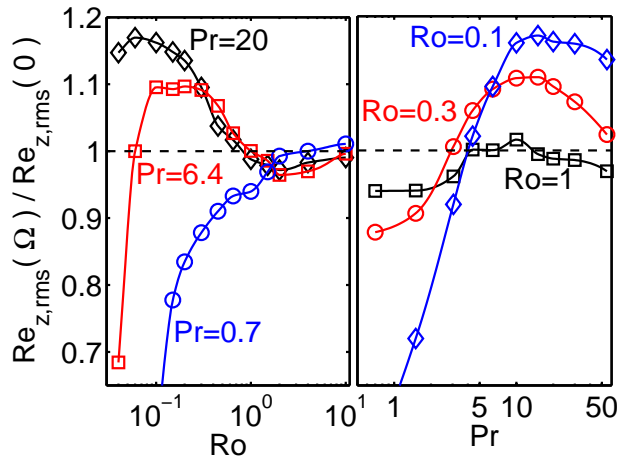}}
\caption{The normalized horizontally averaged rms vertical velocities fluctuation $Re_{z,rms}$ at the edge of the thermal BL for $Ra=1\times 10^8$. Left: Blue circles, red squares, and black diamonds indicate the data for $Pr=0.7$, $Pr=6.4$, and $Pr=20$, respectively. Right: Black squares, red circles, and blue diamonds indicate the data for $Ro=1$, $Ro=0.3$, and $Ro=0.1$, respectively.}
\label{fig:Ekman2}
\end{figure}

Another way to identify the efficiency of Ekman pumping is to look at the horizontally averaged $Re_{z,rms}$  (dimensionless root mean square velocity of the axial velocity fluctuations) value at the height $\lambda_{\theta}^{sl}(r)$ \cite{ste09}. An increase in the value of $Re_{z,rms}$ shows that Ekman pumping becomes important. In figure \ref{fig:Ekman2}b it is shown that Ekman pumping is strongest for moderate $Pr$, because $Re_{z,rms}$ has a maximum around $Pr=10$. Note that the strength of the Ekman pumping measured in this way reflects the measured $Nu$ number enhancement. One can also deduce from figure \ref{fig:Ekman} that the kinetic BL becomes thinner with respect to the thermal BL when the rotation rate is increased, because the temperature difference with the plate becomes smaller. Therefore more hot fluid is sucked into the vertical vortices and this further supports the Ekman pumping effect for moderate rotation rates where the strongest heat transfer enhancement is observed. Furthermore, the maximum in the heat transfer enhancement as shown in figure \ref{fig:Nu_1} occurs at lower $Ro$, i.e. stronger rotation, for higher $Pr$. The breakdown of $Nu$ at low $Ro$ is an effect of the suppression of vertical velocity fluctuations through the strong rotation, which is seen in figure \ref{fig:Ekman2}a. Moreover, in figure \ref{fig:Ekman2}a it is shown that the vertical velocity fluctuations are suppressed at higher $Ro$, i.e. lower rotation rate, when $Pr$ is lower.

To summarize, we studied the $Pr$ number dependence of the heat transport enhancement in rotating RB convection and showed that at a fixed $Ro$ number there is a $Pr$ number for which the heat transfer enhancement reaches a maximum. This is because Ekman pumping, which is responsible for the heat transfer enhancement, becomes less efficient when the $Pr$ number becomes too small or too large. At small $Pr$ numbers the efficiency is limited due to the large thermal diffusivity due to which the heat that is sucked out of the BLs rapidly spreads out in the bulk. At high $Pr$ numbers the temperature of the fluid that is sucked into the vertical vortices near the bottom plate is much lower, because the thermal BL is much thinner than the kinetic BL, and therefore the efficiency of Ekman pumping is lower. Furthermore, the rotation rate for which the heat transfer enhancement reaches its maximum increases with increasing $Pr$, because the suppression of vertical velocity fluctuations only becomes important at larger rotation rates for higher $Pr$.

{\it Acknowledgements:} We thank R.\ Verzicco for providing us with the numerical code. The work is supported by the Foundation for Fundamental Research on Matter (FOM) and the National Computing Facilities (NCF), both sponsored by NWO. The numerical simulations were performed on the Huygens cluster of SARA in Amsterdam.

\bibliographystyle{prsty}


\end{document}